\documentclass[a4paper,twocolumn]{article}
\usepackage{authblk}
\usepackage{siunitx}
\usepackage{booktabs} 
\usepackage{amsmath}
\usepackage{amssymb}
\usepackage{graphicx}
\usepackage{xcolor}
\usepackage[square,numbers]{natbib}
\title{A reduced model for particle calcination for use in DEM/CFD simulations}  

\author[1]{Lucas Reineking}
\author[2]{Torben Bergold}
\author[2]{Enric Illana}
\author[2]{Viktor Scherer}
\author[1]{Martin M{\"o}nnigmann}
\affil[1]{Automatic Control and System Theory, Ruhr University Bochum, Germany}
\affil[2]{Energy Plant Technology, Ruhr University Bochum, Germany}

\begin{document}
\maketitle

\begin{abstract}
    We treat the accurate simulation of the calcination reaction in particles, where the particles are large and, thus, the inner-particle processes must be resolved. 
    Because these processes need to be described with coupled partial differential equations that must be solved numerically, the computation times for a single particle are too high for use in simulations that involve many particles. 
    Simulations of this type arise when the Discrete Element Method (DEM) is combined with Computational Fluid Dynamics (CFD) to investigate industrial systems such as quick lime production in lime shaft kilns.
    
    We show that, based on proper orthogonal decomposition and Galerkin projection, reduced models can be derived for single particles that provide the same spatial and temporal resolution as the original PDE models at a considerably reduced computational cost.
    Replacing the finite-volume particle models with the reduced models results in an overall reduction of the reactor simulation time by about 60\% for the simple example treated here.
\end{abstract}


\message{Column width: \the\columnwidth}
\message{Line width: \the\linewidth}
\message{Text width: \the\textwidth}

\section{Introduction}\label{sec:introduction}
Quick lime ($\text{CaO}$) is an important industrial commodity used for many applications, such as wastewater treatment, flue gas desulfurization or steel production.
Quick lime is produced from lime stone ($\text{CaCO}_3$) by calcination.
The reactors commonly employed for quick lime production are shaft kilns.
Such kilns can reach heights of $20\,\mathrm{m}$, diameters of $4\,\mathrm{m}$ and production capacities of $400\,\mathrm{t/d}$. 
Due to their size, measurements are extremely challenging in these reactors because the moving particle bed is densely packed and high temperatures above $850\,^{\circ}\mathrm{C}$ are present.
Therefore, numerical simulation of these kilns is an attractive alternative. 

Models and numerical tools used for lime shaft kiln simulation vary in complexity.
Many studies are based on one-dimensional heat and mass balances.
For example,~\citet{bes2006dynamic} investigated the operational conditions inside a kiln for different fuels such as lean gas, natural gas and lignite neglecting gradients in the radial direction of the kiln. \citet{shagapov2008theoretical} presented a mathematical model to compute a one-dimensional profile of gas-phase properties and solid temperatures along a kiln considering calcination of limestone and burning of coke.

Other approaches use CFD simulations to resolve the fluid flow. The particles are, however, typically treated in an approximate fashion.
For example,~\citet{mohammadpour2022cfd} carried out CFD simulations of a shaft kiln representing the presence of the particles by a porous medium in the CFD simulations.
Similarly,~\citet{duan2022numerical} employed CFD simulations combined with a shrinking core calcination model on a sub-grid level to investigate the inner state of a shaft kilns \cite{duan2022numerical,duan2022study}.

For a more detailed description of the three-dimensional processes in lime shaft kilns, a method must be used that reflects the correct time and temperature evolution of the limestone bed in the kiln.
Such a method is the Discrete Element Method (DEM). DEM tracks the movement of the individual particles and their reaction. When combined with Computational Fluid Dynamics (CFD), this allows for a detailed description of the thermochemical processes in the kiln.
Studies that combine DEM and CFD in this manner are scarce, however.
A single shaft kiln was investigated with DEM/CFD in
\cite[]{bluhm2010coupled} and \cite[]{krause2015coupled}. 
The same method was applied to a two-shaft regenerative kiln in a subsequent study \cite[]{krause20173d}.
The simulation of a shaft kiln in oxyfuel operation was treated~\cite[]{illana2023oxyfuel}, where combustion takes place in a $\text{O}_2-\text{CO}_2$ atmosphere to allow for $\text{CO}_2$ separation at the kiln gas outlet. 

DEM/CFD simulations for lime shaft kilns are very time-consuming as the number of particles can easily reach more than one million, and particle and gas-phase reaction must be described in addition to particle and fluid flow.
The present paper demonstrates that the models and simulations for the intra-particle calcination processes can be replaced by reduced order models. In contrast to the original PDE models, the reduced models consist of ordinary differential equations (ODEs) that require only a fraction of the computational time, while the error of the reduction can be controlled.  
Several different model reduction methods have been explored for their potential to reduce the overall computation time of DEM/CFD with complex particle models, among them methods based on proper orthogonal decomposition and projection both for particles and flow fields (see, e.g., \cite{Sommer2023}),   
methods based purely on machine learning (see, e.g., \cite{Jendersie2024,Mjalled2023,Mjalled2024}),  
and combinations of these two classes of methods~(see, e.g., \cite{Mjalled2023}).  

We use model reduction based on proper orthogonal decomposition and Galerkin projection (see, e.g., \cite{benner2015survey}). This choice is motivated by their successful application to DEM/CFD simulations of drying processes. These investigations started with single particles~\cite{Scherer2016} and proving they can be used for optimal control~\cite{berner2020controllability} 
and state reconstruction purposes~\cite{berner2020observer}. 
More recently, it has been shown that DEM/CFD simulations for wood chip drying can be accelerated by a factor of about three if the original single particle PDE models are replaced with reduced models~\cite{Reineking2024}. 



In contrast to drying, which can be modeled with coupled diffusion processes, calcination involves reactions and, hence, exponentially depends on temperature as long as transport does not limit the conversion. More specifically, the conversion of limestone to quick lime inside the particles is the challenge. 
The local conversion rate is determined by the local partial pressure of $\text{CO}_2$, which is in turn governed by the local production of $\text{CO}_2$ of the calcination and the transport of $\text{CO}_2$ in the gas phase.
The gas-phase transport in limestone is considerably different from that in quick lime 
because quick lime has got a distinctly higher porosity, which leads to a lower transport resistance. 
Effectively, a reaction front is formed that moves from the particle surface to the particle core.

The intra-particle model for calcination is introduced in section~\ref{sec:model}. Subsequently, the  derivation of the reduced model is explained in section~\ref{sec:reduced-model}. The sample reactor system, and the methods used to simulate it, are described in section~\ref{sec:bulk-setup}.
Reference simulation results obtained with the finite-volume method and the reduced model are presented and compared in section~\ref{sec:results}. 

\section{Particle model}\label{sec:model}

The particle is modeled as a homogeneous isotropic porous medium. The solid phase is composed of limestone ($\text{CaCO}_3$) which reacts to quicklime ($\text{\text{CaO}}$) according to the endothermic calcination reaction 
\begin{equation}
    \text{CaCO}_3(s) \, \rightarrow  \, \text{CaO}(s) \, + \, \text{CO}_2(g) \, + \Delta^R H .
    \label{calc_reaction}
\end{equation}
The reaction progress $R$ is defined as the mass ratio between produced mass of quicklime $m_{\text{CaO}}$ and the maximum mass of quicklime $m_{\text{CaO},\max}$ that can be produced, i.e.,
\begin{equation}
    \label{eq:reaction-progress}
    R = \frac{m_{\text{CaO}}}{m_{\text{CaO},\max}} 
    = \frac{m_{\text{CaO}}}{m_{\text{CaCO}_{3},0}}\frac{M_{\text{CaCO}_3}}{M_{\text{CaO}}}
\end{equation}
where $m_{\text{CaCO}_3,0}$ denotes the initially present mass of limestone,
and $M_{\text{CaCO}_3}$ and $M_{\text{CaO}}$ are the molar masses of $\text{CaCO}_3$ and $\text{CaO}$, respectively.

The gas phase, a mixture of air and $\text{CO}_2$ from~\eqref{eq:reaction-progress} with density $\varrho$ and $\text{CO}_2$ mass fraction $y$, is transported through the pores of the particle. This is modeled with the porosity approach, i.e., the intra-particle pores are not resolved and all quantities are volume-averaged.
The conservation equations for mass and species are stated in terms of density $\rho$ and mass fraction of carbon dioxide $y$. They read 
\begin{subequations}
    \label{eq:calcination-pde}
    \begin{align}
        \frac{\partial}{\partial t} (\varepsilon \varrho)+\nabla \cdot (\varepsilon\varrho \mathbf{v}) &= M_{\text{CO}_2}\dot{n}\label{eq:calcination-conservation-gas} \\
        \frac{\partial}{\partial t} (\varepsilon \varrho y)+\nabla \cdot (\varepsilon\varrho y \mathbf{v}) &= \nabla \cdot (\varepsilon\varrho D_\mathrm{eff} \cdot \nabla y) + M_{\text{CO}_2}\dot{n}\label{eq:calcination-conservation-co2}
    \end{align}
\end{subequations}
with three-dimensional gas-phase velocity $\mathbf{v}$, molar reaction rate $\dot{n}$ of~\eqref{calc_reaction},
and molar mass $M_{\text{CO}_2}$ of $\text{CO}_2$.
The effective diffusivity $D_\mathrm{eff}$ takes the binary diffusion of $\text{CO}_2$ in air and Knudsen diffusion into account.
Details are given in \eqref{eq:effective-diffusivity}~-~\eqref{eq:binary-diffusion} in appendix~\ref{app:closures-continuum-model}.
The intra-particle porosity $\varepsilon$, defined as the fraction of  the volume occupied by the gas phase to the volume occupied by both phases, depends on the reaction progress according to
\begin{equation}
    \label{eq:porosity}
    \varepsilon = R \varepsilon_{\text{CaO}} + (1-R) \varepsilon_{\text{CaCO}_3},
\end{equation}
where $\varepsilon_{\text{CaO}}$ and $\varepsilon_{\text{CaCO}_3}$ denote the porosity of quicklime and limestone, respectively.
Note that the porosity of quicklime is much larger than the porosity of limestone (see Table~\ref{tab:properties}).

The velocity vector $\mathbf{v}$ determines the advection of the gas phase through the porous particle. We calculate it with Darcy's law
\begin{equation}
	\label{eq:darcy}
	\mathbf{v} = - \frac{K}{\mu} \nabla p,
\end{equation}
where the permeability $K$ is a material constant and $\mu$ denotes the fluid dynamic viscosity.
The gas-phase pressure $p$ is calculated according to the ideal gas law
\begin{equation*}
    \varrho R_m T = M p,
\end{equation*}
where $M$ denotes the molar mass of the gas phase.

The reaction rate $\dot{n}$ depends on the local temperature $T$ and concentration of $\text{CO}_2$ $c_{\text{CO}_2}$ according to the Arrhenius equation
\begin{equation*}
    \dot{n} = k_0 A_\mathrm{eff} T \, \exp \bigg(  -\frac{E_A}{R_m T} \bigg) Y_\mathrm{TC} \,(c_{\text{CO}_2} - c_\mathrm{eq}(T))
\end{equation*}
where $k_0$, $A_\mathrm{eff}$, $E_A$, $R_m$ and $Y_\mathrm{TC}$ denote the constant preexponential factor (in m/s), the effective reaction surface (according to the size of the control volume), the activation energy, the universal gas constant and the temperature correction factor, respectively.
The equilibrium concentration $c_\mathrm{eq}(T)$ is calculated according to the ideal gas law 
\begin{equation*}
    c_\mathrm{eq} = \frac{p_\mathrm{eq}(T)}{R_m T}, 
\end{equation*}
where the equilibrium pressure $p_\mathrm{eq}(T)$ can be determined with the empirical correlation~\eqref{eq:equilibrium-pressure} given in appendix~\ref{app:closures-continuum-model}.

The balance equations~\eqref{eq:calcination-pde} need to be combined with an energy balance, which is stated in terms of the temperature $T$. We recall the temperature must be resolved inside the particle, because the particles are large. 
The temperature of the gas and solid phase are assumed to be equal. This approximation is valid because due to the large surface of the interface between the two phases.
The PDE for the conservation of energy for the combined solid-gas phases reads
\begin{equation}
    \label{eq:thermo-pde}
    \frac{\partial}{\partial t} (\rho c_p T) = \nabla \cdot ( \lambda \cdot \nabla T ) + M_{\text{CO}_2}\,\Delta^R H\, \dot{n},
\end{equation}
where the heat capacity $c_p$ and thermal conductivity $\lambda$ are mass-averaged from the individual phases. They depend on the temperature and composition of the gas and the solid phases.
As the particle diameter stays constant during calcination,
the density $\rho$ can be computed from the sum of masses of the phases. The loss of solid mass in the reaction~\eqref{calc_reaction} must be compensated by an increase of porosity.

Fluid and particles exchange heat and mass flux across the particle surface.
Hence, the system of governing equations of~\eqref{eq:calcination-pde} and~\eqref{eq:thermo-pde} is completed by the boundary conditions
\begin{subequations}
    \label{eq:boundary-conditions}
    \begin{align}
        \rho c_p \frac{\partial T}{\partial\mathbf{n}}\vert_{\partial\Omega} &= \alpha (T_\infty -T)\vert_{\partial\Omega} \label{eq:boundary-heatflux}\\
        \varepsilon\varrho D_\mathrm{eff} \frac{\partial y}{\partial\mathbf{n}}\vert_{\partial\Omega} &= \beta\varrho_\infty(y_\infty - y)\vert_{\partial\Omega} \label{eq:boundary-convflux}\\
        \varepsilon\varrho \frac{K}{\mu} \frac{\partial P}{\partial\mathbf{n}}\vert_{\partial\Omega} &= C^* (p_\infty - p)\vert_{\partial\Omega}\label{eq:boundary-advflux}
    \end{align}
\end{subequations}
for the convective heat flux~\eqref{eq:boundary-heatflux}, the convective transport of $\text{CO}_2$~\eqref{eq:boundary-convflux}, and the advective mass flux~\eqref{eq:boundary-advflux},
where $T_\infty$, $\rho_\infty$ and $p_\infty$ are the properties of the fluid from the CFD simulation.  
The heat transfer coefficient, the convective mass transfer coefficient and the advective transfer coefficient are denoted by $\alpha$, $\beta$ and $C^*$, respectively.
All transfer coefficients are computed from the empirical correlations \eqref{eq:heat-mass-transfer-corr}~-~\eqref{eq:pressure-transfer} given in appendix~\ref{app:closures-continuum-model}.

The partial differential equations~\eqref{eq:calcination-pde} and \eqref{eq:thermo-pde} are discretized with a 3D finite volume scheme of tetrahedral cells. The validation of the model and its sensitivity to the spatial discretization are discussed in appendix~\ref{app:validation-continuum-model}. 

\section{Reduced particle model}\label{sec:reduced-model}
The reduced model essentially replaces the PDEs~\eqref{eq:calcination-pde}, \eqref{eq:thermo-pde} and the finite volume solver required to solve them with a set of spatial modes $\varphi^{(\cdot)}$ and time-dependent weighting factors $a^{(\cdot)}(t)$ for these modes. The time-dependent linear combination of the modes $\varphi^{(\cdot)}$ then yields the temperature $T(x, t)$, gas phase density $\rho(x, t)$ and $\text{CO}_2$ mass fraction $y(x, t)$ in the particle as a function of location $x$ and time $t$ according to 
\begin{subequations}
    \label{eq:approximation}
    \begin{align}
        T(x,t) &\approx \bar{T}(x) + \kappa_T \sum_{i=1}^r \varphi^{(T)}_{i}(x) a^{(T)}_{i}(t) \\
        \varrho(x,t) &\approx \bar{\varrho}(x) + \kappa_\varrho \sum_{i=1}^r \varphi^{(\varrho)}_{i}(x) a^{(\varrho)}_{i}(t) \\
        y(x,t)& \approx \bar{y}(x) + \kappa_y \sum_{i=1}^r\varphi^{(y)}_{i}(x) a^{(y)}_{i}(t) 
    \end{align}
\end{subequations}
where $\bar{T}(x)$, $\bar{\rho}(x)$ and $\bar{y}(x)$ are the temporal means and $\kappa_T$, $\kappa_{\rho}$ and $\kappa_y$ are scale factors introduced for numerical reasons. 
The description of the calcination process with~\eqref{eq:approximation} is much more computationally efficient than solving the PDE model~\eqref{eq:calcination-pde}, \eqref{eq:thermo-pde} because the modes $\varphi^{(\cdot)}$ only have to be determined once as a preparation, 
and the coefficients $a^{(\cdot)}$ are governed by ODEs that replace the PDEs~\eqref{eq:calcination-pde}, \eqref{eq:thermo-pde}. 
Moreover, a small number of ODEs, coefficients $a_i^{(\cdot)}$ and modes $\varphi^{(\cdot)}$ suffices to achieve a high precision. This number, which is denoted by $r$ in~\eqref{eq:approximation}, amounts to $r= 5$ for the calcination particle model treated here (see section~\ref{sec:results}).  

We briefly describe how to determine the modes $\varphi^{(\cdot)}_i$ and the ODEs that govern the coefficients $a_i^{(\cdot)}$ in sections~\ref{subsec:modes} and~\ref{subsec:Galerkin}.
These sections are given in order for the paper to be self-contained.

\subsection{Determining the modes}\label{subsec:modes}
The modes $\varphi^{(\cdot)}_{i}(x)$, as well as $T(x, t)$, $\rho(x, t)$ and $y(x, t)$, are denoted as continuous functions of $x$ and $t$ for convenience in~\eqref{eq:approximation}. 
These quantities are approximated on the same grid used in the finite volume solver. Let $\Omega$, $N$ and $x_j$ refer to the spatial domain the particle occupies, the number of cells used to discretize $\Omega$, and the cell center of cell $j$, respectively. Assume the finite volume method provides results at the $M$ time points $t_k= k\Delta t$, $k= 0, \dots, M-1$. 
%
The matrix that holds the results provided by the finite volume solver is called snapshot matrix. For $T(x, t)$ this matrix reads 
\begin{equation*}
    \tilde{T} = \begin{bmatrix}
        \vert & \, & \vert \\
        T(t_0,x_j) & \cdots & T(t_{M-1},x_j) \\
        \vert &\, & \vert  
    \end{bmatrix} \in \mathbb{R}^{N\times M}.
\end{equation*}
The temporal mean $\bar{T}$ required in~\eqref{eq:approximation} 
is the row average of $\tilde{T}$.
The scale factor $\kappa_T$ used in~\eqref{eq:approximation} is chosen to be
\begin{equation*}
    \kappa_T = \max_{x_j,\,t_k} \vert \tilde{T} - \bar{T}\vert.
\end{equation*}
The snapshot matrices, temporal means and scale factors for density $\varrho$ and mass fraction $y$ are calculated analogously.
We calculate the modes $\varphi_i^{(T)}(x)$ 
$$
    \begin{bmatrix}
        \vert\\
        \varphi_i^{(T)}(x_j)\\
        \vert
    \end{bmatrix}
    \in\mathbb {R}^N, \,
    i= 1, \dots, r
$$ 
from the thin singular value decomposition of the scaled meanfree snapshot matrix
\begin{equation*}
    \begin{aligned}
        \frac{1}{\kappa_T}(\tilde{T}-\bar{T}) &= \Phi_T S_T W_T^\intercal \\
        \varphi_i^{(T)}&= \left(
  \Phi_T\right)_{\cdot, i}.
    \end{aligned}
\end{equation*}
where $\Phi_T\in\mathbb{R}^{N\times M}$, $S_T\in\mathbb{R}^{M\times M}$, and $W_T\in\mathbb{R}^{M\times M}$, which implies $\mathrm{rank} (\tilde{T}-\bar{T}) = M$.

The first $r$ columns of $\Phi_T$ are known to be the optimal truncated basis with $r$ elements for the column space of $(\tilde{T}- \bar{T})/\kappa_T$ according to the Eckart-Young theorem~\cite[see Algorithm~8.6.2,][]{golub2013matrix}. 
Consequently, these $r$ columns are the optimal choice for the approximation~\eqref{eq:approximation}. 

In general, the spatial domain $\Omega$ is discretized with finite volume cells of varying volume $V_j$.
Therefore, a volume-weighted scalar product is introduced for functions $f,g$ on the domain $\Omega$
\begin{equation}
    \label{eq:scalar-product}
    \langle f(x), g(x) \rangle = \int_\Omega f(x)\,g(x)\,\mathrm{d}v
    \approx \sum_{j=1}^{N} f(x_j) g(x_j) V_j
\end{equation}
The modes $\varphi^{(\cdot)}(x_j)$,
which are defined for each finite volume cell center $x_j$, can be understood as the spatial discretization of $\varphi^{(\cdot)}(x)$ with a continuous $x\in\Omega$. 
The approximation~\eqref{eq:scalar-product} therefore reads
$\langle \varphi^{(\cdot)}_i, \varphi^{(\cdot)}_j
\rangle
=\langle \varphi^{(\cdot)}_i(x), \varphi^{(\cdot)}_j(x)
\rangle= $
\begin{equation}\label{eq:ScalarProductHelper}
\int_{\Omega} 
\varphi^{(\cdot)}_i(x), \varphi^{(\cdot)}_j(x) dv
\approx 
\sum_{k=1}^N \varphi^{(\cdot)}_i(x_k) \varphi^{(\cdot)}_j(x_k) V_k
\end{equation}
for the modes. We use the left hand side of~\eqref{eq:ScalarProductHelper} whenever convenient.
The scalar product~\eqref{eq:scalar-product} is used to scale the modes $\varphi^{(\cdot)}_{i}$ according to
    \begin{align*}
        \langle\varphi^{(\cdot)}_{i},\varphi^{(\cdot)}_{j}\rangle &= \frac{1}{3} \delta_{ij}
    \end{align*}
where $\delta_{ij}$ is Kronecker's delta. 

\subsection{Galerkin Projection}\label{subsec:Galerkin}
It remains to derive ODEs that govern the coefficients $a_i^{(\cdot)}$ required for the approximation~\eqref{eq:approximation}. 
These ODEs are derived by substituting~\eqref{eq:approximation} into the PDEs~\eqref{eq:calcination-pde} and \eqref{eq:thermo-pde} and calculating the scalar products~\eqref{eq:scalar-product} of the resulting equations with the modes $\varphi_i^{(\cdot)}$. The variant of this procedure used here is called Galerkin projection. 
We demonstrate the Galerkin projection with the mass fraction $y$ 
because the dependence of the projected ODEs appears for this case.

Forming the scalar product~\eqref{eq:scalar-product} of the first $r$ modes $\varphi_i^{(y)}$ with~\eqref{eq:calcination-conservation-co2} yields
\begin{equation}
    \label{eq:pde-y-proj-norms}
    \begin{aligned}
        \langle \varphi^{(y)}_{i}(x), \frac{\partial}{\partial{t}}&(\varepsilon \varrho y)\rangle =
          \: \langle \varphi^{(y)}_{i}(x), \nabla \cdot (D_\mathrm{eff}\varrho \nabla{y} - \varepsilon \varrho y \mathbf{v} ) \rangle \\
        + \: &\langle \varphi^{(y)}_{i}(x), M_{\text{CO}_2} \dot{n} \rangle, \qquad i=1,\ldots,r.
    \end{aligned}
\end{equation}
Note that $r$ linearly independent equations are created.
It is convenient to treat the scalar products in~\eqref{eq:pde-y-proj-norms} separately.
Applying the chain rule to the left hand side of~\eqref{eq:pde-y-proj-norms} yields
\begin{subequations}
\begin{equation}
    \langle \varphi^{(y)}_{i}(x), \frac{\partial}{\partial{t}}(\varepsilon \varrho y)\rangle =
        \sum_{j=1}^{r}
            M^{(yy)}_{ij} \: \dot{a}^{(y)}_{j}(t)
            + M^{(y\varrho)}_{ij}\: \dot{a}^{(\varrho)}_{j}(t)
\end{equation}
with 
    \label{eq:pde-y-proj-lefthandside}
    \begin{align}
        M^{(yy)}_{ij} &= \kappa_y \int_\Omega \varphi^{(y)}_{i}(x) \, \varepsilon \varrho \varphi^{(y)}_{j}(x)\,\mathrm{d}v
         \\
        M^{(y\varrho)}_{ij} &= \kappa_\varrho \int_\Omega \varphi^{(y)}_{i}(x) \, \varepsilon y \varphi^{(\varrho)}_{j}(x)\,\mathrm{d}v ,
    \end{align}
\end{subequations}
where $M^{(yy)}_{ij},M^{(y\varrho)}_{ij}\in\mathbb{R}$.
Using Gauss theorem and substituting the approximations~\eqref{eq:approximation}, the first scalar products on the right handside of~\eqref{eq:pde-y-proj-norms} results in
\begin{equation*}
    \langle \varphi^{(y)}_{i}(x), \nabla \cdot (D_\mathrm{eff}\varrho \nabla{y}
    - \varepsilon \varrho y \mathbf{v} ) \rangle =
    S^{(y)}_i - I^{(y)}_i
\end{equation*}
where 
\begin{subequations}
    \label{eq:pde-y-gauss-applied}
    \begin{align}
        S^{(y)}_i &= \int_{\partial\Omega}  \varphi^{(y)}_{i}(x) \: \big(D_\mathrm{eff}\varrho \nabla{y} - \varepsilon \varrho y \mathbf{v} \big)\cdot \mathbf{n} \,\mathrm{d}a\label{eq:y-surface-fluxes}\\
        I^{(y)}_i &= \int_{\Omega} \nabla \varphi^{(y)}_{i}(x) \cdot \big(D_\mathrm{eff}\varrho \nabla{y} -\varepsilon \varrho y \mathbf{v} \big) \,\mathrm{d}v\label{eq:y-inner-fluxes}
    \end{align}
\end{subequations}
can be interpreted as the surface fluxes $S^{(y)}\in\mathbb{R}^R$ and the innner fluxes $I^{(y)}\in\mathbb{R}$, respectively. This interpretation is advantageous for two reasons.
First, due to the surface integral in~\eqref{eq:y-surface-fluxes} it is possible to explicitly incorporate the boundary conditions~\eqref{eq:boundary-convflux} and~\eqref{eq:boundary-advflux} into the reduced model. This yields the surface fluxes 
\begin{equation}
    S^{(y)}_i = \int_{\partial\Omega}  \varphi^{(y)}_{i}(x) \: \big(
        \beta\varrho_\infty(y_\infty - y)+y C^*(p_\infty-p)
        \big) \,\mathrm{d}a
\end{equation}
Second, only first derivatives appear in~\eqref{eq:y-inner-fluxes}.
The remaining scalar product in~\eqref{eq:pde-y-proj-norms} reads
\begin{equation}
    \label{eq:y-inner-source}
    \langle \varphi^{(y)}_{i}(x), M_{\text{CO}_2} \dot{n} \rangle = 
    \int_\Omega \varphi^{(y)}_{i}(x)  M_{\text{CO}_2} \dot{n} \,\mathrm{d}v = J^{(y)}_i,
\end{equation}
which defines the $i$th element of the inner source terms $J^{(y)}\in\mathbb{R}^r$.

Substituting~\eqref{eq:pde-y-proj-lefthandside}-\eqref{eq:y-inner-source} 
into~\eqref{eq:pde-y-proj-norms}
yields the ODE for the mass fraction $y$
\begin{equation}\label{eq:CoupledODEs_a_y}
    M^{(yy)} \dot{a}^{(y)}(t) + M^{(y\varrho)} \dot{a}^{(\varrho)}(t) = S^{(y)} - I^{(y)} + J^{(y)}.
\end{equation}
Note that both $\dot{a}^{(y)}$ and $\dot{a}^{(\rho)}$ appear in~\eqref{eq:CoupledODEs_a_y}. In this sense, the ODEs for $a^{(y)}$ and $a^{(\rho)}$ are coupled. 
 
The procedure that yields~\eqref{eq:CoupledODEs_a_y} is analogously applied to the PDEs~\eqref{eq:calcination-conservation-co2} and \eqref{eq:thermo-pde}. Combining the results for the Galerkin projections of all three PDEs yields 
\begin{equation}
    \mathbf{M} \mathbf{\dot{a}} = \mathbf{S} - \mathbf{I} + \mathbf{J}
\end{equation}
where
\begin{equation}
    \mathbf{M} = \begin{bmatrix}
        M^{(TT)} & 0 & 0 \\
        0 & M^{(\varrho\varrho)} & 0 \\
        0 & M^{(y\varrho)} & M^{(yy)}
    \end{bmatrix} \in \mathbb{R}^{3r\times 3r}
\end{equation}
and\begin{equation}
    \label{eq:rom-ode-vectors}
    \mathbf{\dot{a}} = \begin{bmatrix}
        \dot{a}^{(T)} \\ \dot{a}^{(\varrho)} \\ \dot{a}^{(y)}
    \end{bmatrix},
    \:
    \mathbf{S} = \begin{bmatrix}
        S^{(T)} \\ S^{(\varrho)} \\ S^{(y)}
    \end{bmatrix},
    \:
    \mathbf{I} = \begin{bmatrix}
        I^{(T)} \\ I^{(\varrho)} \\ I^{(y)}
    \end{bmatrix},
    \:
    \mathbf{J} = \begin{bmatrix}
        J^{(T)} \\ J^{(\varrho)} \\ J^{(y)}
    \end{bmatrix}
    \in\mathbb{R}^{3r}.
\end{equation}
The results of the reduced model for a single particle with constant boundary conditions are given in Figure~\ref{fig:redunced-single-particle-model} in appendix~\ref{app:rom-single-particle}.

\section{Sample system}\label{sec:bulk-setup}
\begin{figure}
	\centering
    \includegraphics[width=0.6\linewidth]{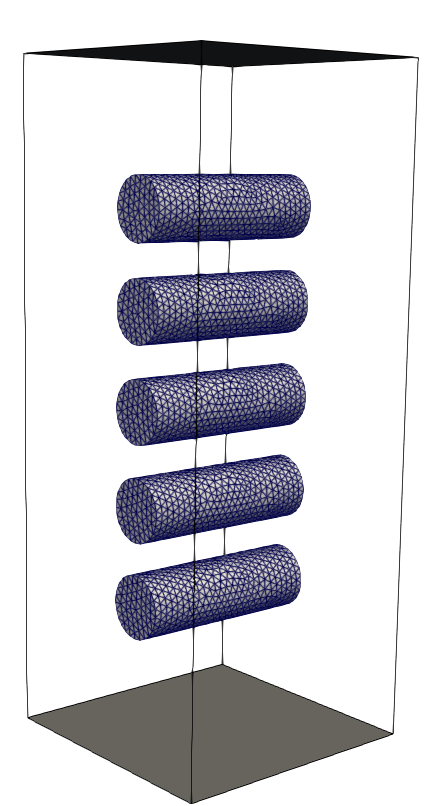}
	\caption{System consisting of 5 limestone particles distributed uniformly over the height}
	\label{fig:bulk}
\end{figure}

We consider a rectangular reactor with a simple arrangement of particles as a test case.  
The domain of interest measures
$50\,\mathrm{mm}$ in width, $50\,\mathrm{mm}$ in depth, and $125\,\mathrm{mm}$ in height. 
Five cylindrical particles with a diameter of $12\,\mathrm{mm}$ and a length of $36\,\mathrm{mm}$ are arranged vertically with  $6\,\mathrm{mm}$ separation distance between them ($18\,\mathrm{mm}$ axis to axis, see Figure~\ref{fig:bulk}). 
The cylinders are discretized by 8000 tetrahedral control volumes each. 
The particles are assumed to initially consist of pure limestone with an uniform initial temperature of 900 $^\circ \text{C}$.  
The material properties for limestone and quick lime are given in Table~\ref{tab:properties}.

\begin{table*}[htb]%
	\centering
	\caption{Properties for lime and quicklime according to \cite{hills1968} }
	\label{tab:properties}
	\begin{tabular}{llcS[table-format=5.2]S[table-format=4.2]}
		\toprule
		%
        property &~& unit & \multicolumn{1}{c}{$\text{CaCO}_3$} & \multicolumn{1}{c}{CaO} \\
		\midrule
		density	& $\varrho_\text{j}$ & $\left[ \frac{\text{kg}}{\text{m}^3}\right]$ & 2812.05 & 3310 \\
		heat capacity & $c_\text{p,j}$ & $\left[\frac{\text{J}}{\text{kgK}}\right]$ & 836.8 & 753.1\\
		thermal conductivity & $\lambda_\text{j}$ & $\left[\frac{\text{W}}{\text{mK}}\right]$ & 2.26 & 0.07\\
		porosity & $\epsilon_\text{j}$ & $[-]$ & 0.04 & 0.543\\
		specific surface & $S_\text{j}$ & $\left[\frac{\text{m}^2}{\text{kg}}\right]$ & 16000 & 7000\\
		\bottomrule
	\end{tabular}%
\end{table*}
Air enters the reactor from the bottom ($z = 0$) with a superficial velocity $v_{\text{in}}=0.5\,\mathrm{m/s}$, mass flow $\dot{m}_{\text{in}}=0.375\,\mathrm{g/s}$ and temperature $T_{\text{in}}=900\,\mathrm{^\circ C}$,  
and leaves the reactor at the top end ($z = 0.125\,\mathrm{m}$) where the ambient pressure is set to $p_{\text{out}}=101325\,\mathrm{Pa}$. The particle Reynolds number amounts to $\mathrm{Re}_0 = 228$ based on an equivalent sphere. 
 
		%
		%
		%
		%
		
		%
		%


DEM and CFD simulations are coupled with the porous medium approach, also called averaged volume method (AVM).
In this approach, particles are represented in the flow field by a porosity field, which implies the detailed particle shape is not considered in the CFD simulations.
The porous media and the CFD are coupled with source terms for momentum, energy and species from DEM to CFD. Conversely, the CFD solution provides far-field boundary conditions, which are uniform along the particle surface, to the porous medium simulation. 
Since the purpose of the paper is to investigate the acceleration that can be achieved with reduced models for the particles, we used a one-dimensional model for the gas phase for simplicity. 
The domain is spatially discretized with 7 control volumes in the $z$-direction ($\delta z=18\,\mathrm{mm}$). Each particle is coupled only with the control volume in which its center of gravity is located. A zero gradient condition is imposed at the side boundaries for all quantities.
All simulations are carried out with an inhouse DEM code. OpenFOAM version 20.12 \cite{openfoam} 
is used for simulating mass, species and energy transport in the fluid phase.\footnote{
All simulations were conducted with 
a single core AMD(R) EPYC(R) 7443 CPU @ 2.85GHz and 256 GB RAM under Ubuntu 20.04.5 LTS
}

Because of the small number of particles, an immersed boundary method (IBM)~\citep[see, e.g.,][]{GORGES20241} would also be applicable to accurately describe particle shape in the flow field. However, we deliberately have refrained from IBM as it will be too costly for large-scale simulations of technical reactors with several hundred thousands of particles, which are the motivation of our work.

\section{Results}\label{sec:results}
\begin{figure}
	\centering
         \includegraphics[width=\columnwidth]{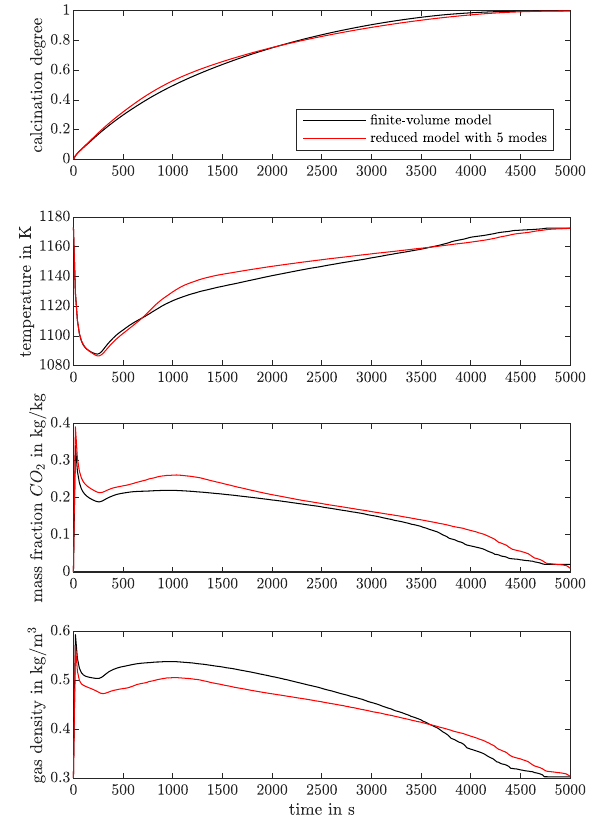}
	\caption{Spatially averaged (over all particles) calcination degree, temperature, $\text{CO}_2$ mass fraction and gas-phase density inside the particle over time.}
 \label{fig:integrated_5p_comp}
\end{figure}

Figure~\ref{fig:integrated_5p_comp} depicts the temporal evolution of the calcination degree, temperature, $\text{CO}_2$ mass fraction and gas-phase density. All diagrams in Figure~\ref{fig:integrated_5p_comp} show the spatial average over the five particles.
As the particles have an initial uniform temperature of 900 K, the $\text{CO}_2$ production driven by kinetics is initially very high in the whole particle, leading to a peak in the $\text{CO}_2$ mass fraction. In parallel, the particle temperature drops fast due to the endothermic calcination reaction initially present in the whole particle. The fast $\text{CO}_2$ release at the beginning also leads to an initial peak in density, as the $\text{CO}_2$ produced is initially stored in the particle and not enough $\text{CO}_2$ can leave the particle, i.e., an overpressure builds up in the particle. The $\text{CO}_2$ mass fraction decreases very fast after the initial peak since the advective mass flow, which is proportional to the intra-particle pressure, is large leading to a high driving force.
As soon as this initial phase has settled to a minimum temperature and an equilibrated pressure at around 250 s, the calcination progresses steadily inwards. This is accompanied by an increase in particle temperature as heat is transferred by convection from the gas to the particle and, in addition,  the outer control volumes become fully calcinated, i.e., the heat sink is then absent. 
From 250 seconds until approximately 1000 seconds, the $\text{CO}_2$ mass fraction and gas density increase again with moderate gradients as the particle temperature is still at a low level, i.e., the reaction rate is slow. Nevertheless, $\text{CO}_2$ production is larger than the amount of $\text{CO}_2$ being diffused and advected outside the particles, leading again to a storage effect of $\text{CO}_2$ in the particle. After 1000 seconds, due to the reduction in the reaction area of the reaction front, diffusion and advection become predominant, and the $\text{CO}_2$ mass fraction as well as density decrease to the values of the surrounding gas. 
In general, the reduced-order model with only 5 modes and 5 ODEs is able to reproduce the results obtained with the finite DEM/CFD simulation in a satisfactory manner.
This holds for the calcination degree in particular, which is the main parameter defining quick lime quality. 

\begin{figure}
	\centering
         \includegraphics[width=.5\columnwidth]{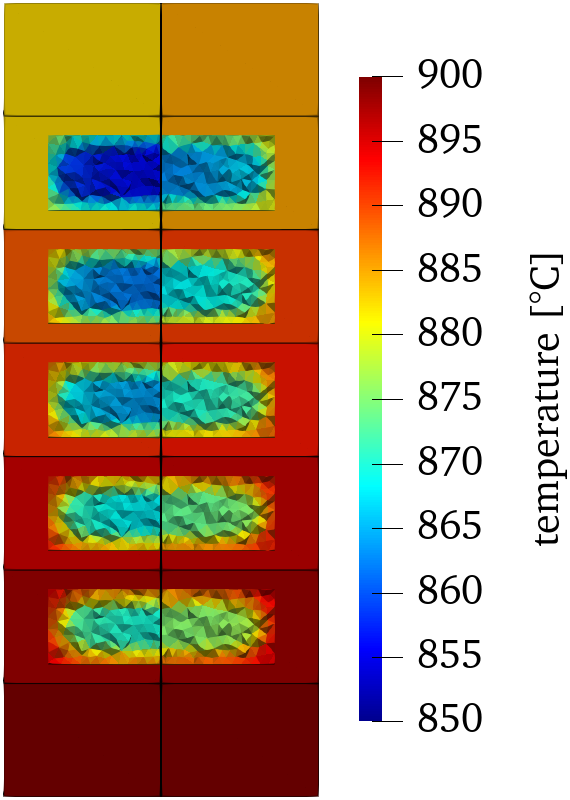}
	\caption{Temperature in the center cross section of the reaction volume at time $t= 2500\text{s}$. DEM/CFD results shown to the left and reduced-order model results shown to the right of the black vertical line. } 
	\label{fig:result_T}
\end{figure}

\begin{figure}
	\centering
         \includegraphics[width=.5\columnwidth]{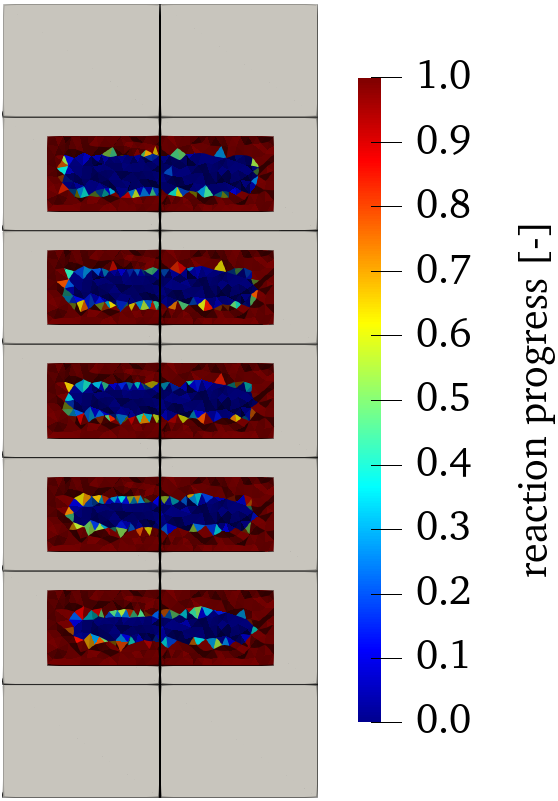}
	\caption{Reaction progress in the center cross section of the reaction volume at time $t= 2500\text{s}$. DEM/CFD results shown to the left and reduced-order model results shown to the right of the black vertical line.  } 
	\label{fig:result_reaction_progress}
\end{figure}

\begin{figure}
	\centering
         \includegraphics[width=.5\columnwidth]{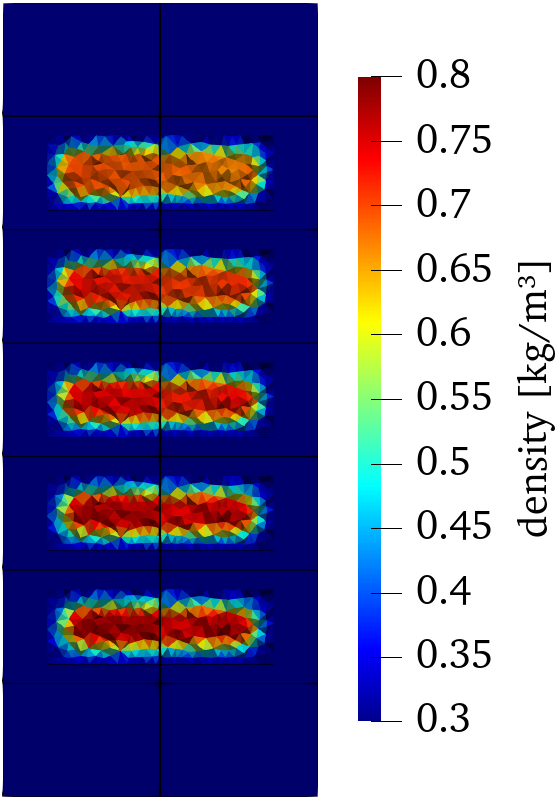}
	\caption{Gas phase density at the center cross section of the reaction volume at time $t= 2500\text{s}$. DEM/CFD results shown to the left and reduced-order model results shown to the right of the black vertical line.}
	\label{fig:result_rho}
\end{figure}

\begin{figure}
	\centering
         \includegraphics[width=.5\columnwidth]{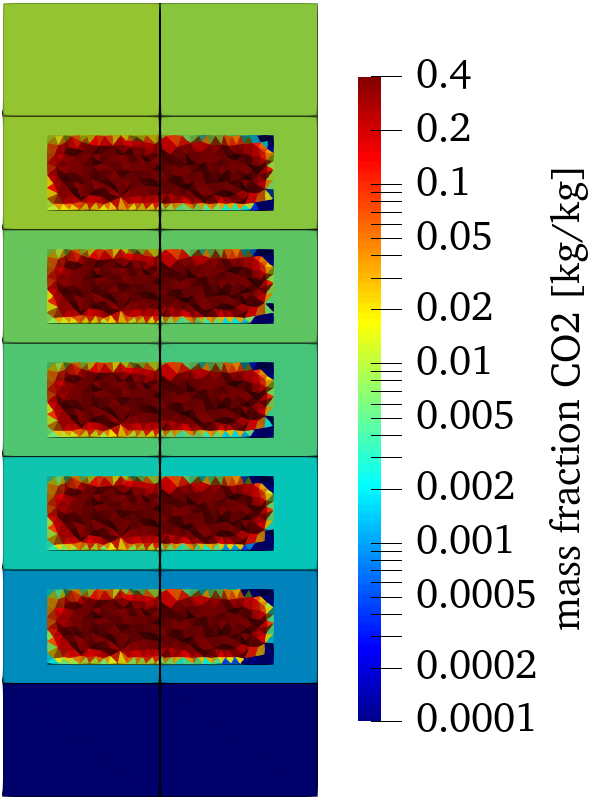}
	\caption{$\text{CO}_2$ concentration at the center cross section of the reaction volume at time $t= 2500\text{s}$. DEM/CFD results shown to the left and reduced-order model results shown to the right of the black vertical line.}
	\label{fig:result_CO2}
\end{figure}

Figures~\ref{fig:result_T}--\ref{fig:result_CO2} show vertical cross-sectional views through the center plane of the reactor. Black vertical lines divide the diagrams into left and right parts that show the results of the DEM/CFD simulation and the results obtained with the reduced-order model, respectively. 

The gas phase temperature is depicted in Figure~\ref{fig:result_T}. It is obvious that the gas-phase temperature decreases from the gas inlet to the outlet. This is due to the fact that the gas phase transfers heat to the particles along its path to the outlet to cover the endothermic calcination reaction. 
For each of the particles, the temperature in the center is lower than at the surface. In the outer layers, calcination progress is already 1 (see Figure~\ref{fig:result_reaction_progress})  and the heat transferred from the gas phase is no longer required to cover the heat for calcination but available for particle heating. 
The mean temperature of the particle further downstream is at a lower level, as the progress of calcination is lower there, i.e., more heat is still needed for the endothermic calcination reaction.  
The reduced-order model results in slightly higher temperatures in the interiors of the particles. This deviation, which amounts to a few Kelvin, is consistent with the temperature values in Figure~\ref{fig:integrated_5p_comp} at time $t=2500\text{s}$. 

Figure~\ref{fig:result_reaction_progress} shows the reaction progress as defined in~\eqref{eq:reaction-progress} for the particles. 
It resembles the intra-particle $\text{CO}_2$ mass fraction (see Figure~\ref{fig:result_CO2}), however, is not identical to it.
Whereas the reaction progress describes local conversion of $\text{CaCO}_3$ to $\text{CaO}$, the $\text{CO}_2$ mass fraction only gives an information of the local $\text{CO}_2$ concentration in the particle. This implies, for example, a cell can have a non-zero $\text{CO}_2$ mass fraction even if no reaction takes place in it, because of diffusion of  $\text{CO}_2$ from an neighboring cell. The results obtained with the DEM/CFD simulation and the reduced-order model are in good agreement in Figure~\ref{fig:result_reaction_progress}. 

The gas phase density is depicted in Figure~\ref{fig:result_rho}. 
The density is determined by the balance of $\text{CO}_2$ released from the limestone and the transport of the $\text{CO}_2$ away from the location it is produced at. When the transport is slower than $\text{CO}_2$ production, the local density (and the pressure) increases. This is reflected in Figure~\ref{fig:result_rho}. In the center of the particle, where the calcination reaction is active, the pressure is larger than in outer layers, where reaction progress is already 1, i.e., no $\text{CO}_2$ is produced any longer and  transport to the particle surface takes place. 
It is evident from Figure~\ref{fig:result_rho} that 
the results obtained with the reduced-order model are in very good agreement for the $\text{CO}_2$ concentration. 

Figure~\ref{fig:result_CO2} shows that the $\text{CO}_2$ released from the limestone is accumulated in the gas phase, i.e., the gas phase $\text{CO}_2$ concentration increases in the direction of the fluid flow. 
The $\text{CO}_2$ mass fraction in the particle is lower at the reactor entrance than further downstream for two reasons. 
First, the accumulated $\text{CO}_2$ in the fluid further downstream hinders the mass transfer of $\text{CO}_2$ from the particle core to the particle surface, i.e., more $\text{CO}_2$ remains in the particles that are located closer to the gas outlet of the reactor. 
Second, as the calcination reaction has progressed further for the particles at the entrance, the porosity of these particles is already large due to the presence of the porous  CaO in the outer layers of the particle (see~\eqref{eq:porosity}), i.e., resistance for mass transport to the surface is lower for the particles at the reactor entrance. 
In general, the $\text{CO}_2$ concentration in the center of the particle is larger than in outer particle layers. 

The reduced-order model results in lower $\text{CO}_2$ mass fractions than the DEM/CFD simulation. Specifically, these deviations appear at the surface. When integrated over the surface, these discrepancies are compensated. As a consequence, the sum of $\text{CO}_2$ source terms are similar for the finite-volume and the reduced-order model results, which results in very similar $\text{CO}_2$ mass fractions across the entire fluid phase. 


For the simulation results shown here, the reduced model from section~\ref{sec:reduced-model} required approximately 40\% of the computational effort needed by the finite volume method from section~\ref{sec:model}.


\addcontentsline{toc}{section}{Acknowledgments}
\section*{Acknowledgments}
This work was funded by the Deutsche Forschungsgemeinschaft (DFG, German Research Foundation) - Project-ID 422037413 - TRR 287. Gef{\"o}rdert durch die Deutsche Forschungsgemeinschaft (DFG) - Projektnummer 422037413 - TRR 287.

\appendix
\section{Closure terms for the calcination model}
\label{app:closures-continuum-model}
The continuum model stated in section~\ref{sec:model} requires several closures which are given for completeness. 
The equilibrium pressure $p_\mathrm{eq}(T)$ of the calcination reaction only depends on the temperature $T$. It is given by~\cite{silcox1989mathematical}
\begin{equation}
    \label{eq:equilibrium-pressure}
\begin{split}
    p_\mathrm{eq}&(T) = 101325\,\,\exp\big(17.74 \\ &- 0.00108T + 0.332 \ln(T) - \frac{22020}{T} \big).
\end{split}  
\end{equation}
The enthalpy of reaction $\Delta^R H(T)$ of the reaction~\eqref{calc_reaction} can be derived from Kirchhoff's law for the reactants. This yields
\begin{equation}
    \label{eq:enthalpy-reaction}
\begin{split}
\Delta^R H(T) & =\Delta H_\mathrm{form}(\theta)|_{\theta=298\text{K}}+\int_{298K}^T c_{p,{CO_2}}(\theta)\,  M_{CO_2} \\
& + c_{p,CaO}  M_{CaO}-c_{p,{CaCO_3}}  M_{CaCO_3} \text{d}\theta,
\end{split}
\end{equation}
where $c_{p,i}$ and $M_i$ denote the heat capacity and the molar mass of specie $i$, respectively.
The formation enthalpy $\Delta H_\mathrm{form}$ is a constant of the reaction.

The effective diffusivity $D_\mathrm{eff}$ is computed from the binary diffusivity of $\text{CO}_2$ in air $D_b$ and the Knudsen diffusivities $D_{\mathrm{Kn},CaO}$ and $D_{\mathrm{Kn},CaCO_3}$
\begin{equation}
    \label{eq:effective-diffusivity}
    D_\mathrm{eff} = \frac{\varepsilon}{\tau^2} \cdot \bigg(\frac{R}{\frac{1}{D_b}+\frac{1}{D_{\mathrm{Kn},CaO}}}
        + \frac{1 - R}{\frac{1}{D_b}+\frac{1}{D_{\mathrm{Kn},CaCO_3}}}\bigg),
\end{equation}
where $\tau$ denotes tortosity. 
The Knudsen diffusivities are calculated from
\begin{equation}
    \label{eq:knudsen-diffusion}
    D_{\mathrm{Kn},i} = \frac{2}{3} r_{\mathrm{Pore},i}\sqrt{\frac{8R_m T}{\pi M_{CO_2}}}
    \qquad i=\text{CaO}, \text{CaCO}_3.
\end{equation}
The binary diffusivity of $\text{CO}_2$ in air is given by~\cite{fuller1966new}
\begin{equation}
    \label{eq:binary-diffusion}
    D_b = \frac{1.343}{p} \,\left(\frac{T}{273}\right)^{1.75},
    \quad p\:\mathrm{in\:Pa},\:T\:\mathrm{in\:K}
\end{equation}

Further, the heat and mass transfer coefficients $\alpha$ and $\beta$,  are calculated from
\begin{subequations}
    \label{eq:heat-mass-transfer-corr}
    \begin{align}
        \alpha &= \mathrm{Nu}\,\,\frac{\lambda_\infty}{L}\\
        \beta  &= \mathrm{Sh}\,\,\frac{D_b}{L},
    \end{align}
\end{subequations}
where the characteristic length $L$ is set to the square root of the surface area of the facet of the finite volume tetrahedron.
Nusselt number $\mathrm{Nu}$ and Sherwood number $\mathrm{Sh}$ are approximated by the correlation
\begin{subequations}
    \begin{align}
        \mathrm{Nu} &= 0.664\,\,\mathrm{Re}^{1/2}\mathrm{Pr}^{1/3}\\
        \mathrm{Sh} &= 0.664\,\,\mathrm{Re}^{1/2}\mathrm{Sc}^{1/3}.
    \end{align}
\end{subequations}
The Schmidt number $\mathrm{Sc}$ is calculated from properties of the surrounding fluid
\begin{subequations}
    \label{eq:sherwood-prandtl}
    \begin{align}
        \mathrm{Pr} &= 0.71 \\
        \mathrm{Sc} &= \frac{\eta_\infty}{\varrho_\infty D_b}
    \end{align}
\end{subequations}
The advective mass transfer coefficient is obtained from Darcy's law. This results in
\begin{equation}
    \label{eq:pressure-transfer}
    C^* = \frac{K\varrho_\infty}{\mu_\infty L^*}
\end{equation}
with the characteristic length $L^*$ set to the radius of the volume equivalent sphere of the local tetrahedron.

\section{Validation of the calcination model}
\label{app:validation-continuum-model}
\begin{figure}
	\includegraphics{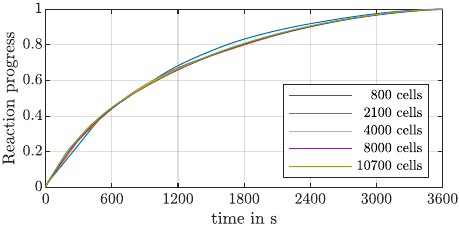}
	\caption{Impact of discretization on the calcination degree of the cylindrical particles}
	\label{fig:mesh_independence}
\end{figure}

We varied the number of tetrahedral cells used in the finite-volume model from $800$ to $10,700$ cells to determine an appropriate resolution. 
We evaluated the calcination degree as a function of the cell number for this purpose. 
The calcination degree is defined as the ratio of $\text{CO}_2$ mass released from a lime particle to the total initial mass of $\text{CO}_2$ bound in $\text{CaCO}_3$ in the particle. It can be calculated from
\begin{equation}
    \label{eq:calcination-degree}
    D_\mathrm{calcination} = \frac{\int_0^t \dot{m}_{\text{CO}_2,\mathrm{\text{out}}}\,\mathrm{d}\tau}{\int_\Omega \rho_{\text{CaCO}_3}-\rho_{\text{CaO}}\,\mathrm{d}v}
\end{equation}
Figure~\ref{fig:mesh_independence} depicts the calcination degree as a function of time for a gas-phase temperature of 900 \textdegree C and $\text{CO}_2$ mass fraction in the gas-phase of 0 $\frac{kg}{kg}$.
The results are very similar for a number of cells above 2100. 

\begin{figure}
	\includegraphics{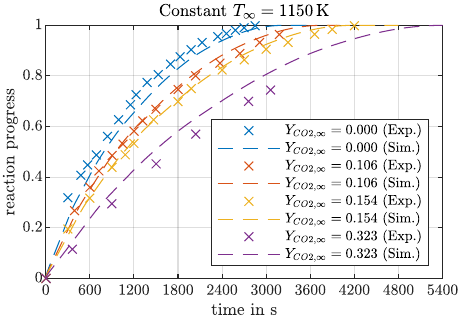}
	\caption{Comparison of simulation results obtained with the finite-volume model to experimental data~\cite{hills1968}, part I. Calcination degree for a sphere for various $CO_2$ mass fractions.}
	\label{fig:validation-const-T}
\end{figure}

\begin{figure}
	\includegraphics{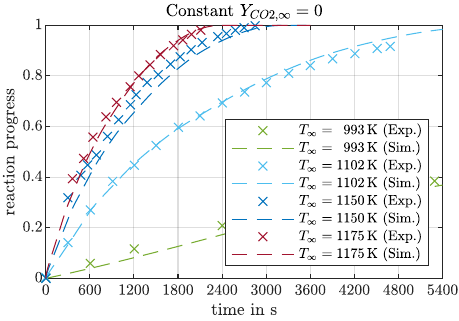}
	\caption{Comparison of simulation results obtained with the finite-volume model to experimental data~\cite{hills1968}, part II.  Calcination degree for a sphere for various temperatures.}
	\label{fig:validation-const-Y}
\end{figure}

As a further validation, we compared finite-volume simulation results obtained with 1700 cells to experimental data for a single spherical particle (diameter $10\text{mm}$) from the literature~\cite{hills1968}. 
This comparison is shown in  Figures~\ref{fig:validation-const-T} and~\ref{fig:validation-const-Y}, where the calcination degree is plotted as a function of time for different ambient $\text{CO}_2$ concentrations at constant gas-phase temperature of $1150\,\mathrm{K}$ (top), and for a set of gas-phase temperatures at zero ambient $\text{CO}_2$ concentration (bottom).
As expected, a larger ambient $\text{CO}_2$ concentration (Figure~\ref{fig:validation-const-T}) slows down calcination reaction, and an increase in ambient gas-phase temperature accelerates calcination (Figure~\ref{fig:validation-const-Y}).
The model is able to reproduce these effects with reasonable
accuracy. 

\section{Validation of the reduced-order model}
\label{app:rom-single-particle}
Before using it in the simulations summarized in section~\ref{sec:bulk-setup},
we validated the single-particle reduced-order model
with $r= 5$ modes and ODEs by comparing it to finite-volume simulations with 8000 cells. 
The fluid properties were set to  $T_\infty=1175\,\text{K}$, $y_\infty=0$ and $p_\infty=101325\,\text{Pa}$ and kept constant over time for this validation. 
The initial particle temperature was $1175\,\text{K}$ and the gas phase was assumed to be in the equilibrium state, such that there is no calcination reaction in the initial state.

\begin{figure}
    \includegraphics{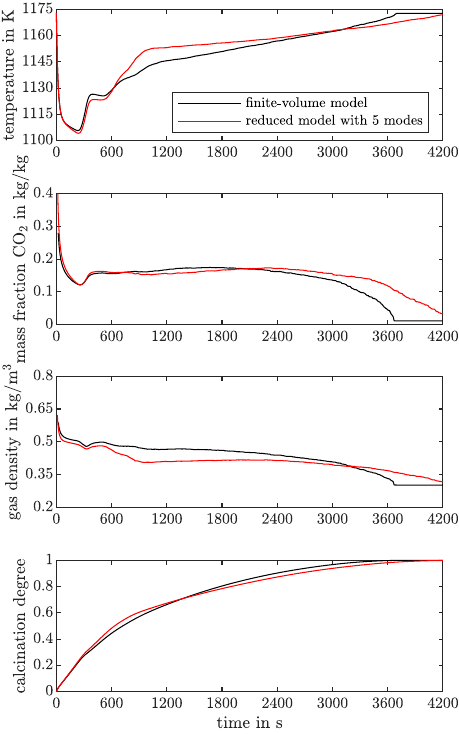}
    \caption{Spatially averaged (over all particles) calcination degree, temperature, $\text{CO}_2$ mass fraction
and gas-phase density in the particle over time used for validation. \label{fig:redunced-single-particle-model}}
\end{figure}

The results for the particle temperature, $\text{CO}_2$ mass density and calcination degree averaged over all 800 cells are shown in Figure~\ref{fig:redunced-single-particle-model}.  
For the first approximately $60\,\text{s}$, $\text{CO}_2$ is driven out of the particle because of the difference of the $\text{CO}_2$ concentration in the particle and the fluid.
The calcination reaction then starts 
and the mean temperature of the particle sharply declines.
A minimum in temperature is reached at approximately $300\,\text{s}$.
Subsequently, the heat due convective heat transfer from the fluid to the particle exceeds the heat consumed by the calcination reaction. 
In this first period, the calcination reaction is mostly located at the surface of the particle.
We stress that the reduced model reproduces this first period particularly well. This is evident from Figure~\ref{fig:redunced-single-particle-model}.

At around $400\,\text{s}$ a second period begins which is marked by an approximately constant average mass fraction of $\text{CO}_2$ inside the particle.
The calcination reaction takes place at a moving reaction front that travels from the particle surface inwards.
The reaction front can be observed also in Figure~\ref{fig:result_T}, where a sharp gradient of the reaction progress can be seen.
While this moving reaction front persists, an equilibrium state is reached in the sense that approximately the same amount of $\text{CO}_2$ is released at the reaction front as is transferred to the fluid at the particle surface.
This indicates that the reaction is limited by the transport of $\text{CO}_2$  in the particle.
We note that the reduced model shows differences in the prediction of the temperature and the gas density. As a result, the reaction front travels effectively slower in the reduced model, resulting in longer time of the total calcination reaction of the particle.
The slower movement of the reaction front in the reduced model is reflected in a slower increase of the calcination degree in the reduced model between approximately $1000\,\text{s}$ and $3000\,\text{s}$ compared to the finite-volume model.

The last period is reached in the finite-volume model at around $2700\,\text{s}$ and with the reduced model at around $3600\,\text{s}$.
At this time the moving reaction front reached the core and the state variables converge to the respective fluid properties.

\bibliography{references}

\end{document}